\documentclass[final]{IEEEtran}
\usepackage{amsthm,amssymb,graphicx,multirow,amsmath,color,amsfonts,physics}
\usepackage[update,prepend]{epstopdf}
\usepackage[noadjust]{cite}
\usepackage{tikz}
\usepackage{bbm} 
\usepackage{pdfpages}
\usepackage{balance}
\usepackage{multirow}
\usepackage{comment}
\usepackage{subfigure}

\allowdisplaybreaks 

\begin{document}
\def\nba{{\mathbf{a}}}
\def\nbb{{\mathbf{b}}}
\def\nbc{{\mathbf{c}}}
\def\nbd{{\mathbf{d}}}
\def\nbe{{\mathbf{e}}}
\def\nbf{{\mathbf{f}}}
\def\nbg{{\mathbf{g}}}
\def\nbh{{\mathbf{h}}}
\def\nbi{{\mathbf{i}}}
\def\nbj{{\mathbf{j}}}
\def\nbk{{\mathbf{k}}}
\def\nbl{{\mathbf{l}}}
\def\nbm{{\mathbf{m}}}
\def\nbn{{\mathbf{n}}}
\def\nbo{{\mathbf{o}}}
\def\nbp{{\mathbf{p}}}
\def\nbq{{\mathbf{q}}}
\def\nbr{{\mathbf{r}}}
\def\nbs{{\mathbf{s}}}
\def\nbt{{\mathbf{t}}}
\def\nbu{{\mathbf{u}}}
\def\nbv{{\mathbf{v}}}
\def\nbw{{\mathbf{w}}}
\def\nbx{{\mathbf{x}}}
\def\nby{{\mathbf{y}}}
\def\nbz{{\mathbf{z}}}
\def\nb0{{\mathbf{0}}}
\def\nb1{{\mathbf{1}}}

\def\nbA{{\mathbf{A}}}
\def\nbB{{\mathbf{B}}}
\def\nbC{{\mathbf{C}}}
\def\nbD{{\mathbf{D}}}
\def\nbE{{\mathbf{E}}}
\def\nbF{{\mathbf{F}}}
\def\nbG{{\mathbf{G}}}
\def\nbH{{\mathbf{H}}}
\def\nbI{{\mathbf{I}}}
\def\nbJ{{\mathbf{J}}}
\def\nbK{{\mathbf{K}}}
\def\nbL{{\mathbf{L}}}
\def\nbM{{\mathbf{M}}}
\def\nbN{{\mathbf{N}}}
\def\nbO{{\mathbf{O}}}
\def\nbP{{\mathbf{P}}}
\def\nbQ{{\mathbf{Q}}}
\def\nbR{{\mathbf{R}}}
\def\nbS{{\mathbf{S}}}
\def\nbT{{\mathbf{T}}}
\def\nbU{{\mathbf{U}}}
\def\nbV{{\mathbf{V}}}
\def\nbW{{\mathbf{W}}}
\def\nbX{{\mathbf{X}}}
\def\nbY{{\mathbf{Y}}}
\def\nbZ{{\mathbf{Z}}}

\def\ncalA{{\mathcal{A}}}
\def\ncalB{{\mathcal{B}}}
\def\ncalC{{\mathcal{C}}}
\def\ncalD{{\mathcal{D}}}
\def\ncalE{{\mathcal{E}}}
\def\ncalF{{\mathcal{F}}}
\def\ncalG{{\mathcal{G}}}
\def\ncalH{{\mathcal{H}}}
\def\ncalI{{\mathcal{I}}}
\def\ncalJ{{\mathcal{J}}}
\def\ncalK{{\mathcal{K}}}
\def\ncalL{{\mathcal{L}}}
\def\ncalM{{\mathcal{M}}}
\def\ncalN{{\mathcal{N}}}
\def\ncalO{{\mathcal{O}}}
\def\ncalP{{\mathcal{P}}}
\def\ncalQ{{\mathcal{Q}}}
\def\ncalR{{\mathcal{R}}}
\def\ncalS{{\mathcal{S}}}
\def\ncalT{{\mathcal{T}}}
\def\ncalU{{\mathcal{U}}}
\def\ncalV{{\mathcal{V}}}
\def\ncalW{{\mathcal{W}}}
\def\ncalX{{\mathcal{X}}}
\def\ncalY{{\mathcal{Y}}}
\def\ncalZ{{\mathcal{Z}}}

\def\nbbA{{\mathbb{A}}}
\def\nbbB{{\mathbb{B}}}
\def\nbbC{{\mathbb{C}}}
\def\nbbD{{\mathbb{D}}}
\def\nbbE{{\mathbb{E}}}
\def\nbbF{{\mathbb{F}}}
\def\nbbG{{\mathbb{G}}}
\def\nbbH{{\mathbb{H}}}
\def\nbbI{{\mathbb{I}}}
\def\nbbJ{{\mathbb{J}}}
\def\nbbK{{\mathbb{K}}}
\def\nbbL{{\mathbb{L}}}
\def\nbbM{{\mathbb{M}}}
\def\nbbN{{\mathbb{N}}}
\def\nbbO{{\mathbb{O}}}
\def\nbbP{{\mathbb{P}}}
\def\nbbQ{{\mathbb{Q}}}
\def\nbbR{{\mathbb{R}}}
\def\nbbS{{\mathbb{S}}}
\def\nbbT{{\mathbb{T}}}
\def\nbbU{{\mathbb{U}}}
\def\nbbV{{\mathbb{V}}}
\def\nbbW{{\mathbb{W}}}
\def\nbbX{{\mathbb{X}}}
\def\nbbY{{\mathbb{Y}}}
\def\nbbZ{{\mathbb{Z}}}

\def\nfrakR{{\mathfrak{R}}}

\def\nrma{{\rm a}}
\def\nrmb{{\rm b}}
\def\nrmc{{\rm c}}
\def\nrmd{{\rm d}}
\def\nrme{{\rm e}}
\def\nrmf{{\rm f}}
\def\nrmg{{\rm g}}
\def\nrmh{{\rm h}}
\def\nrmi{{\rm i}}
\def\nrmj{{\rm j}}
\def\nrmk{{\rm k}}
\def\nrml{{\rm l}}
\def\nrmm{{\rm m}}
\def\nrmn{{\rm n}}
\def\nrmo{{\rm o}}
\def\nrmp{{\rm p}}
\def\nrmq{{\rm q}}
\def\nrmr{{\rm r}}
\def\nrms{{\rm s}}
\def\nrmt{{\rm t}}
\def\nrmu{{\rm u}}
\def\nrmv{{\rm v}}
\def\nrmw{{\rm w}}
\def\nrmx{{\rm x}}
\def\nrmy{{\rm y}}
\def\nrmz{{\rm z}}

\def\nbydef{:=}
\def\nborel{\ncalB(\nbbR)}
\def\nboreld{\ncalB(\nbbR^d)}
\def\sinc{{\rm sinc}}

\newtheorem{lemma}{Lemma}
\newtheorem{thm}{Theorem}
\newtheorem{definition}{Definition}
\newtheorem{ndef}{Definition}
\newtheorem{nrem}{Remark}
\newtheorem{theorem}{Theorem}
\newtheorem{prop}{Proposition}
\newtheorem{cor}{Corollary}
\newtheorem{example}{Example}
\newtheorem{remark}{Remark}
\newtheorem{assumption}{Assumption}
	

\newcommand{\ceil}[1]{\lceil #1\rceil}
\def\argmin{\operatorname{arg~min}}
\def\argmax{\operatorname{arg~max}}
\def\figref#1{Fig.\,\ref{#1}}%
\def\E{\mathbb{E}}
\def\EE{\mathbb{E}^{!o}}
\def\P{\mathbb{P}}
\def\pc{\mathtt{P_c}}
\def\rc{\mathtt{R_c}}   
\def\p{p}

\def\V{\operatorname{Var}}
\def\erfc{\operatorname{erfc}}
\def\erf{\operatorname{erf}}
\def\opt{\mathrm{opt}}
\def\R{\mathbb{R}}
\def\Z{\mathbb{Z}}

\def\LL{\mathcal{L}^{!o}}
\def\var{\operatorname{var}}
\def\supp{\operatorname{supp}}

\def\N{\sigma^2}
\def\T{\beta}							
\def\sinr{\mathtt{SINR}}			
\def\snr{\mathtt{SNR}}
\def\sir{\mathtt{SIR}}
\def\ase{\mathtt{ASE}}
\def\se{\mathtt{SE}}

\def\calN{\mathcal{N}}
\def\FE{\mathcal{F}}
\def\calA{\mathcal{A}}
\def\calK{\mathcal{K}}
\def\calT{\mathcal{T}}
\def\calB{\mathcal{B}}
\def\calE{\mathcal{E}}
\def\calP{\mathcal{P}}
\def\calL{\mathcal{L}}


\def\l{\ell}
\newcommand{\fad}[2]{\ensuremath{\mathtt{h}_{#1}[#2]}}
\newcommand{\h}[1]{\ensuremath{\mathtt{h}_{#1}}}

\newcommand{\err}[1]{\ensuremath{\operatorname{Err}(\eta,#1)}}
\newcommand{\FD}[1]{\ensuremath{|\mathcal{F}_{#1}|}}



\def\Bx{{\mathcal{B}}^x}
\def\Bxx{{\mathcal{B}}^{x_0}}
\def\jx{y}
\def\m{(\bar{n}-1)}
\def\mm{\bar{n}-1}
\def\Nx{{\mathcal{N}}^x}
\def\Nxo{{\mathcal{N}}^{x_0}}
\def\wj{w_{jx_0}}
\def\uij{u_{jx}}
 \def\yj{y}
 \def\yjx{y}
 \def\zjx{z_x}
 \def \tx {y_0}
 \def \htx {h_0}

\def\rx{z_{1}}
\def\ry{z_{2}}

\def\Rx{Z_{1}}
\def\Ry{Z_{2}}

\def \hyxx {h_{y_{x_0}}}
\def \hyx {h_{y_x}}

\def\nbb1{\mathbbm{1}}
\def\xi{x_i}
\def\xj{x_j}
\def\xx{x_0}
\def\yk{y_k}
\def\yy{y_0}
\def\ie{{\em i.e. }}
\def\eg{{\em e.g. }}
\def\iid{{\em i.i.d. }}
\def\avg{\rm avg}

\def\rmnuma{\rm\uppercase\expandafter{\romannumeral1}}
\def\rmnumb{\rm\uppercase\expandafter{\romannumeral2}}
\def\rmnumc{\rm\uppercase\expandafter{\romannumeral3}}
\def\rmnumd{\rm\uppercase\expandafter{\romannumeral4}}
\def\rmnume{\rm\uppercase\expandafter{\romannumeral5}}
\def\rmnumf{\rm\uppercase\expandafter{\romannumeral6}}
\pagenumbering{gobble}
\graphicspath{{./Figures/}}
\title{RIS-enabled Integrated Access and Relay:\\ Empowering Collaboration among BSs\\ }
\author{
 Hao Lin,~\IEEEmembership{Graduate Student Member,~IEEE}, Mustafa A. Kishk,~\IEEEmembership{Member,~IEEE}\\ and Mohamed-Slim Alouini,~\IEEEmembership{Fellow,~IEEE}
\thanks{Hao Lin is with the Electrical and Computer Engineering Program, CEMSE
Division, King Abdullah University of Science and Technology (KAUST),
Thuwal 23955-6900, Kingdom Saudi Arabia (e-mail: hao.lin.std@gmail.com).\\
\indent Mustafa A. Kishk is with the Department of Electronic Engineering,
Maynooth University, Maynooth, W23 F2H6 Ireland (e-mail:
mustafa.kishk@mu.ie).\\
\indent Mohamed-Slim Alouini is with the CEMSE Division, King Abdullah
University of Science and Technology (KAUST), Thuwal 23955-6900,
Kingdom Saudi Arabia (e-mail: slim.alouini@kaust.edu.sa).}
}

\maketitle
\begin{abstract}
The increasing number of Internet of Things (IoT) devices and applications leads to severe access congestion in conventional base station (BS) networks. Meanwhile, the low transmit power of IoT devices requires larger diversity gains from the system design. Therefore, low-cost traffic management and signal enhancement mechanisms become essential. Reconfigurable intelligent surfaces (RISs) are increasingly significant due to their flexibility and efficiency. Using reflection, refraction, and amplification, RISs can act as innovative relay nodes to overcome spatial limitations caused by blockages, and improve the coverage range of existing infrastructure. However, more extensive applications of RISs remain to be explored. In this paper, we propose to deploy co-sited RISs on BSs and use the integrated access and relay (IAR) architecture to enable wave-domain task-offloading between BSs, without requiring extra infrastructure or incurring extra decoding latency. Considering the fixed and dynamic sub-carrier schemes, we show the performance improvement in cell-free networks and cellular networks. In addition, we investigate the effect of RIS elements allocated to each IoT device in both cell-free and cellular networks. Finally, we discuss the opportunities and challenges of IAR BSs for future communications design. 
\end{abstract}
\section*{Introduction} \label{sec:intro}
Massive Internet of Things (IoT) deployments introduce new connectivity challenges \cite{9369324}. For instance, multiple IoT devices have to share the same time   and frequency resources, leading to severe access competition. In dense IoT scenarios, BSs are unable to provide satisfactory quality of service to all nearby devices, resulting in significant interference and access congestion. In contrast, in sparse IoT areas, spectral resources are often underutilized. Moreover, many dynamic IoT applications further aggravate traffic load fluctuations. However, redeploying BSs has a very low return on investment (RoI) and cannot adapt to changes in traffic. Therefore, it is necessary to find efficient solutions that make full use of infrastructure and spectrum resources.

\indent Reconfigurable intelligent surfaces (RISs) are emerging due to their potential to enhance the energy efficiency, coverage probability, and capacity of wireless networks, by intelligently reconfiguring the wireless propagation environment \cite{basar2019wireless,liu2021reconfigurable,8741198}. Various kinds of RISs have been proposed to provide reflection, refraction, and amplification  services actively or passively \cite{ni2024single}. Without complex amplify-and-forward (AF) or decode-and-forward (DF) operations, RISs can act as innovative relay nodes (RNs). RISs can be installed on blockages to realize full-space coverage with a lower power consumption and hardware cost than traditional RNs. Through the energy harvesting technique, multifunctional RISs can eliminate the dependence on batteries and the power grid, enabling them to be self-sufficient and support long-term operation \cite{10225701}. Therefore, RISs are considered a cost-effective and energy-efficient solution to empower smart radio environments \cite{renzo2019smart}.


In \cite{9784946}, authors proposed the concept of co-sited RISs, which are deployed in the buildings near the BS to improve the signal and service quality. To address the traffic offloading challenges of massive IoT devices, we propose a cost-effective solution by deploying co-sited RISs at BSs. Specifically, the co-sited RISs enable the redirection of IoT connection requests from existing direct line-of-sight (dLoS) links to indirect line-of-sight (iLoS) BSs and reduces the energy cost of devices searching for available spectrum resources. This cooperative architecture is referred to as integrated access and relay (IAR). In this paper, we systematically investigate the key advantages and operational mechanisms of the proposed IAR architecture.

\section*{Integrated Access and Relay}
\subsection{Operational Mechanism of IAR}
In this paper, we propose to install the multifunctional RISs introduced in \cite{ni2024single,10225701} on BSs, which can achieve omnidirectional signal redirection and amplification. Each BS's antenna and RIS units are managed and optimized by the same control plane. Note that when all RIS components are inactive, the IAR network can be reduced to a conventional network, and no additional signaling overhead will be introduced. When the RIS components are active, additional signaling overhead is incurred, while the IoT devices benefit from diversity gains provided by reflections. In this manuscript, we focus on improving the signal decoding performance from an SINR perspective, therefore, we assume that the signaling overhead and additional latency introduced by the RIS configuration are acceptable when building the link between the IoT device and iLoS BS. The switching frequency of positive-intrinsic-negative (PIN) diodes can be up to 5 megahertz (MHz), which means that the switching time of RIS elements can be 0.2 microseconds ($\mu$s) and is well-suited for the mobile applications with time-varying channels and IoT applications. After RIS reconfiguration, IoT devices can establish a reliable and stable connection with the iLoS BSs. Once a viable physical link is formed via the RIS-assisted path, the IoT device and the iLoS BS further perform synchronization over the established link, enabling subsequent data transmission. In the proposed IAR system, the dLoS BSs only reflect the signal and do not affect the priority of the service. This ensures the fairness of all IoT devices based on the actual signal arrival time.

To reduce the RIS reconfiguration overhead and design complexity, when an IoT device is redirected to potential iLoS BSs via the co-sited RIS elements of its dLoS BS, only one of the pre-defined RIS sub-arrays with a fixed number of RIS elements is used. Therefore, the amplification gain of the signal strength is controllable and predictable. Since the phase reconfiguration of this RIS subarray is optimized for this target device, the signals from other devices are reflected to other directions rather than being coherently combined at the serving iLoS BS. When an IoT device has been served stably, the relative RISs do not further require frequent reconfiguration. The phase changes of RIS elements are triggered by: (i) rapid device movement, (ii) sudden severe shadowing effects on line-of-sight (LoS) links between base stations, (iii) BS node failures, and (iv) system information updates. In the IAR systems, random reflections of signals from other BSs and devices bring the implementation challenges. Therefore, the phase control of RIS elements should maximize the amplification gain and directionality of the serving signal while minimizing interference from other devices on the target reflected signal. The phases of different RIS sub-arrays can be adjusted centrally or independently according to the algorithm complexity. 

\begin{figure*}[ht]
    \centering
    \includegraphics[width=0.8\linewidth]{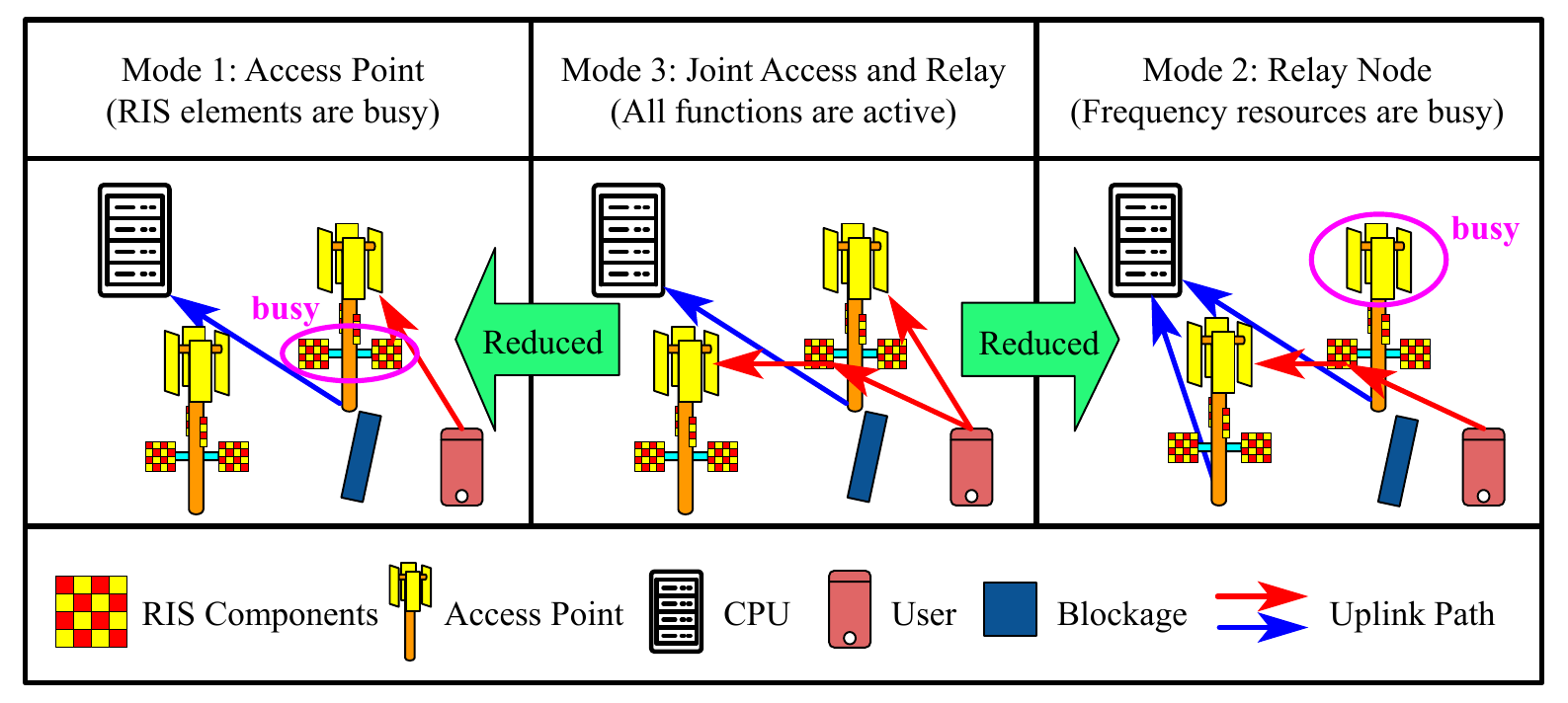}
    \caption{Different modes of IAR BSs, Mode 3 can be reduced to Mode 1 or Mode 2 in different scenarios.}
    \label{fig:mode}
\end{figure*}

As shown in Fig. \ref{fig:mode}, each IAR BS can perform different functions in both cellular and cell-free networks. For example, in cellular networks, wireless devices access the network using only one sub-carrier from one BS in the entire network. When the closest dLoS BSs have idle sub-carriers, wireless devices choose to directly access the antenna of the dLoS BS. This dLoS IAR BS operates as an access node. When the access function of this dLoS BS is busy, it will use one RIS sub-array to reflect signals to a nearby BS and generate an iLoS path. In this case, the dLoS BS operates as an relay node. Especially, in cell-free networks, devices will attempt to access not only the dLoS BS but also the iLoS BSs through the RIS sub-array of the dLoS BS, to have better coverage performance and stable connections even in areas with a large number of IoT devices. Therefore, the dLoS BS operates as an IAR node for this service. When RIS elements or frequency resources on the IAR node are busy or broken, Mode 3 will be reduced to Mode 1 or Mode 2.

Note that, if the load information of the BS network cannot be updated timely, IoT devices need to frequently send requests to different BSs via dLoS and iLoS paths, resulting in significant signaling overhead for both IoT devices and the IAR BS network, where the worst-case has a matching complexity of $\mathcal{O}(N_{\mathrm{BS}} \log N_{\mathrm{BS}})$. However, with periodic load information sharing, feasible iLoS BS candidates can be listed so that the matching complexity can be reduced to $\mathcal{O}(N_{\mathrm{BS}})$, even $\mathcal{O}(1)$. Therefore, in the considered system, we assume that BSs can share or periodically update their load information, where the signaling overhead is manageable. The coordination period between BSs $T_c$ needs to be designed appropriately to adapt to different scenarios. Considering the signaling overhead from BS coordination, $T_c$ should not be set too small. However, in situations with high environmental uncertainty or high mobility of IoT devices, it must be set to a smaller value. Otherwise, once the load information becomes outdated, IoT devices may be unable to connect to the selected target iLoS BS, requiring a longer time to re-match to an available iLoS BS and incurring additional energy costs. Compared to traditional inter-cell cooperation schemes, the BSs equipped with co-sited RISs can further share the deployment location, main reflection direction, operating mode, and scheduled time interval of each RIS sub-array with its neighboring BSs, thereby supporting more accurate channel estimation, power control and inter-cell interference coordination. BS coordination can be implemented in either centralized or distributed ways. Centralized coordination is suitable for small-scale networks or systems with sufficient computational resources, while distributed coordination is more practical for large-scale deployments or scenarios with limited computation capabilities. 

\subsection{Advantages of IAR}
According to the operational mechanism of IAR solutions, we summarize their advantages for future terrestrial networks as follows:
\begin{itemize}
    \item \textbf{Quick deployment and compatibility:} IAR architecture can be quickly implemented by integrating RISs into the BSs. Operators do not have to consider the cost of modifying unauthorized buildings. In addition, the IAR function only reshapes the signal propagation direction without affecting the 5G/6G air interface and framework.
    \item \textbf{Strong system scalability:} By sharing the timely operational status between BSs, RIS elements on each IAR BS can rapidly correct phases and reflect the signals from devices to other BSs with available resource blocks. Such an online algorithm can help ensure that every device receives a service, rather than being rejected. In a large-scale network, off-line routing tables can be designed to reduce the system complexity and control overhead. IAR BS network can pre-determine the access links for potential IoT devices by predicting the traffic information. When some IoT devices fail or new IoT devices connect, the relative IAR BSs should reconfigure the RIS components or device access schemes. When an IAR BS fails or a new IAR BS is established, nearby BSs should update their redirection candidates and optimize the link access schemes for nearby device.
    \item \textbf{Traffic offloading:} For conventional cellular networks, BSs directly serve the devices. However, when the IoT distribution is uneven, some BSs face heavy access loads, while others remain idle. The spectrum and computation resources of the entire network cannot be fully utilized. By embedding such an IAR function into each BS, neighboring BSs can coordinate to form a wireless mesh topology and serve more devices. For example, in a smart city scenario with dense sensor deployments, BSs in hotspot regions can directly serve nearby devices, while redirecting the access requests of remaining devices to the BSs with less load, ensuring most of devices can be connected. 
    \item \textbf{Diversity enhancement:} For conventional cell-free networks, devices can achieve the same QoS through multiple BSs \cite{papazafeiropoulos2020performance}. However, in suburban areas, devices without multiple neighbor dLoS BSs cannot obtain such cell-free service. Using IAR BSs, the iLoS paths can provide higher diversity gain and coverage performance. At the same time, the utilization of the IAR technique also increases the potential burden for the entire network.
\end{itemize}

In simple load-aware BS cooperation without IAR function, after an initial access attempt to its neighbor dLoS BS, devices need to search for other available BSs, but the device-to-BS channel conditions are still unknown. The proposed IAR framework can quickly provides controllable physical links between the IoT device and its iLoS BS, and also provides service enhancement using the gain from signal concentrating. Compared with RIS-assisted relay systems, the IAR solution can significantly reduce the system complexity while providing flexible traffic redirection. However, IAR and traditional RIS-based relay solutions can be used together to reshape the smart radio environment. In massive IoT applications, we consider the narrow-band transmission, and the RIS reflections are reasonably approximated as frequency-flat. In broadband transmission, the RIS reflection amplification model needs adjustment. Nevertheless, IAR solution still provides signal gain and available LoS links for the devices. Therefore, besides IoT applications, IAR solution can also be extended to broadband networks.

\section*{Use cases of IAR BSs} \label{usecases}

\begin{figure}[ht]
    \centering
    \includegraphics[width=1\linewidth]{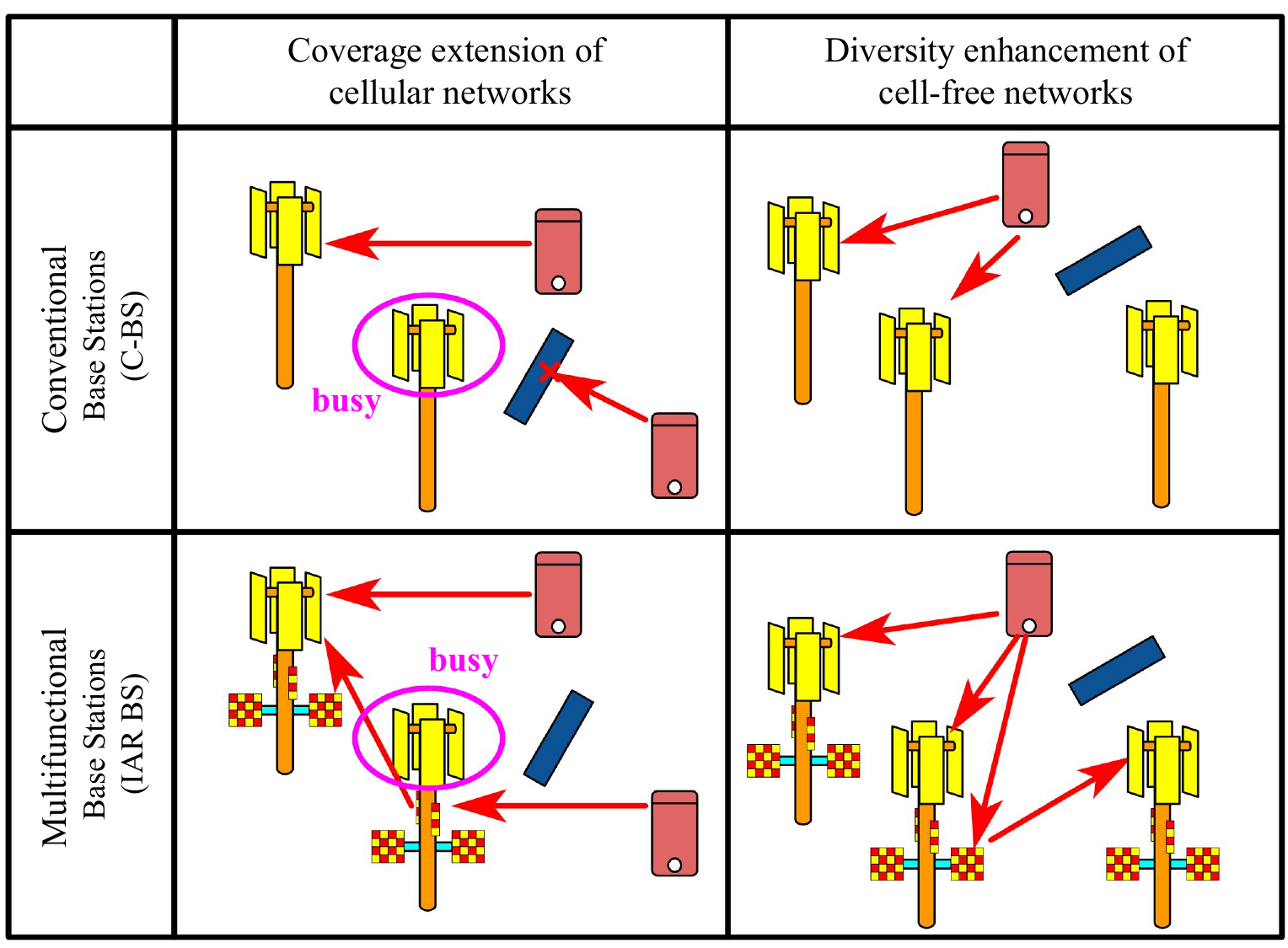}
    \caption{The advantage of IAR BSs in cellular networks and cell-free networks.}
    \label{fig:advantages}
\end{figure}
In this paper, the proposed IAR architecture can provide opportunities for the cooperation between BSs, in both cellular and cell-free networks. However, the use of IAR BSs also leads to a more complex radio environment. In this manuscript, we let $M$ denote the number of effective RIS elements in each RIS sub-array, which determines the achievable signal power gain. Increasing the number of relay services per BS reduces the reflection gain per link, while the signal amplification via RISs intensifies inter-device interference. Therefore, these competing effects jointly result in an optimal value of $M$. Due to the massive number of IoT devices, although the frequency band can be divided into multiple sub-carriers, the sharing of each sub-carrier is still necessary. Therefore, we assume that each BS has $N_r$ frequency resource blocks. To verify whether the IAR technique can achieve performance improvements in different scenarios, we consider the following two sub-carrier management schemes in this paper:
\begin{itemize}
    \item \textbf{Fixed Sub-carrier Scheme:} Each BS has $N_{r}$ frequency resource blocks in total. Each device only operates in its fixed resource block and tries to find its closest dLoS BS to access. If this resource block of the dLoS BS is busy, this dLoS BS uses its $M$ of RIS elements to reflect the signal to its own neighbor LoS BS with available resource block. If the device cannot obtain the service through the closest dLoS BS, it searches for the second closest BS and repeats the process.
    \item \textbf{Dynamic Sub-carrier Scheme:} Each BS has $N_{r}$ frequency resource blocks in total. Each device selects its closest BS whose resource blocks are open to access. The device switches to access the resource block with the least number of serving devices. If all resource blocks on this BS are busy, the closest dLoS BS use its RISs to reflect the signal to its closest LoS BS. If the device cannot obtain the service through this, it searches for the second closest BS and repeats the process.
\end{itemize}

It is worth noting that, the fixed sub-carrier scheme does not require IoT devices to switch sub-carrier frequencies. But when there are many IoT devices, they need to search for farther BSs to connect. The dynamic sub-carrier scheme ensures that more devices can connect to their nearest dLoS BSs, but requires them to dynamically adjust their sub-carrier frequency. In both fixed or dynamic sub-carrier schemes, each resource block can be used to serve multiple devices or one device at the same time. We summarize the characteristics of these two access strategies as below:
\begin{itemize}
    \item \textbf{Multi-Device Access:} Each resource block of each BS can serve multiple devices at the same time. The signals from different devices interfere with each other. The maximum number of devices that access the same resource block is $N_m$. The access request from other devices on the same frequency will be denied when the resource block already serves $N_m$ devices. Such a strategy cannot ignore the impact of interference from other devices, but each resource block can serve as many devices as possible when the decoding SINR threshold is low.
    \item \textbf{Single-Device Access:} Each resource block of each BS serves only one device and denies service requests from any other devices. The single-device access strategy can minimize the influence of interference because each device has an independent frequency band. However, some devices cannot obtain the service in the dense area.
    
\end{itemize}

For both cellular and cell-free networks, IAR BSs can improve coverage performance through the RIS-based relay function. To prove this, we discuss the following two examples:
\begin{itemize}
    \item \textbf{Cellular network with one-hop relay:} Each device tries to find its closest dLoS BS to access. If failed, the closest dLoS BS reflects the signal to its neighbor LoS BS whose resource block is open to access. If both of these paths are not available, the device will find its second closest dLoS BS and repeat the same process.
    \item \textbf{Cell-free network with one-hop relay:} Each device tries to find its closest dLoS BS to access. Regardless of whether the dLoS BS is busy, it attempts to utilize its idle RIS components to reflect the signal to its neighbor dLoS BS that is not busy. Only when such two paths are both unavailable, the device will try to find the second closest dLoS BS and repeat the process.
\end{itemize}

Considering different access strategies, we design Monte Carlo experiments to prove the performance improvement through IAR technique in cellular networks and cell-free networks.

\section*{Proof of Concept and Simulation Results} \label{sec:sim}
We conduct 10,000 Monte Carlo simulations in a $\rm 1\,km\times 1\,km$ suburban area, where blockages (BKs), IoT devices and BSs are randomly distributed in this area. All devices have the same transmit signal power, where the signal propagation suffers from the path loss shown in \cite{kishk2020exploiting} and Nakagami-m fading. We model the blockages following the distribution introduced in \cite{al2014optimal}. The parameters are defined in Table \ref{tab:TableOfNotations} and their values are also given. In Figs. 3–5, we compare the coverage probabilities of the proposed IAR solutions with those of conventional BS (C-BS) networks. In the C-BS networks, IoT devices continuously search for synchronization signals from their closest available dLoS BSs. Differently, IAR-based cellular and cell-free networks operate using the scheme described in the previous section. In both kinds of networks, load information update can help IoT devices find the target BS more quickly, but the decoding performance regarding SINR will not be affected. Therefore, the experimental results of C-BSs are independent of the number of RIS elements per sub-array, represented as a straight line parallel to the x-axis in the figures.

\begin{table}[ht]\caption{Parameters in Simulation}
\centering
    \begin{tabular}{ {l} | {c} | {c} }
    \hline
        \hline
    \textbf{Parameter} & \textbf{Notation} & \textbf{Value} \\ \hline
    The ratio of built-up land area to the total area & $\alpha_{BK}$ & $0.1$\\ \hline
    The number of blockages (BKs) per ${\rm km^2}$ & $\beta_{BK}$ & $750$\\ \hline
    The scale parameter of BKs' heights distribution & $\gamma_{BK}$ & 8\\ \hline
    The number of IoT devices & $N_{D}$ & $200$\\ \hline
    The height of IoT devices & $h_{D}$ & $1\,{\rm m}$\\ \hline
    The number of base stations (BSs) & $N_{BS}$ & $15$\\ \hline
    The height of BSs & $h_{BS}$ & $15\,{\rm m}$\\ \hline
    The number of resource blocks in each BS & $N_{r}$ & $20$\\ \hline
    The uplink power of each IoT device & $P$ & $\rm 0.1\,W$\\ \hline
    The total number of RIS elements on each BS & $M_{tot}$ & $512$\\ \hline
    The path loss exponent & $\alpha$ & $\rm 3$\\ \hline
    The Nakagami-$m$ fading parameters & $(m,\Omega)$ & $(2,1)$\\ \hline
    The noise power & $N_0$ & $10^{-8}$\\ \hline
    The SINR threshold & $\rm SINR_{th}$ & $0.1$\\ \hline
     \hline
    \end{tabular}
\label{tab:TableOfNotations}
\end{table}

\subsection{IAR-empowered cell-free network}
\begin{figure}[ht]
    \centering
    \includegraphics[width=1\linewidth]{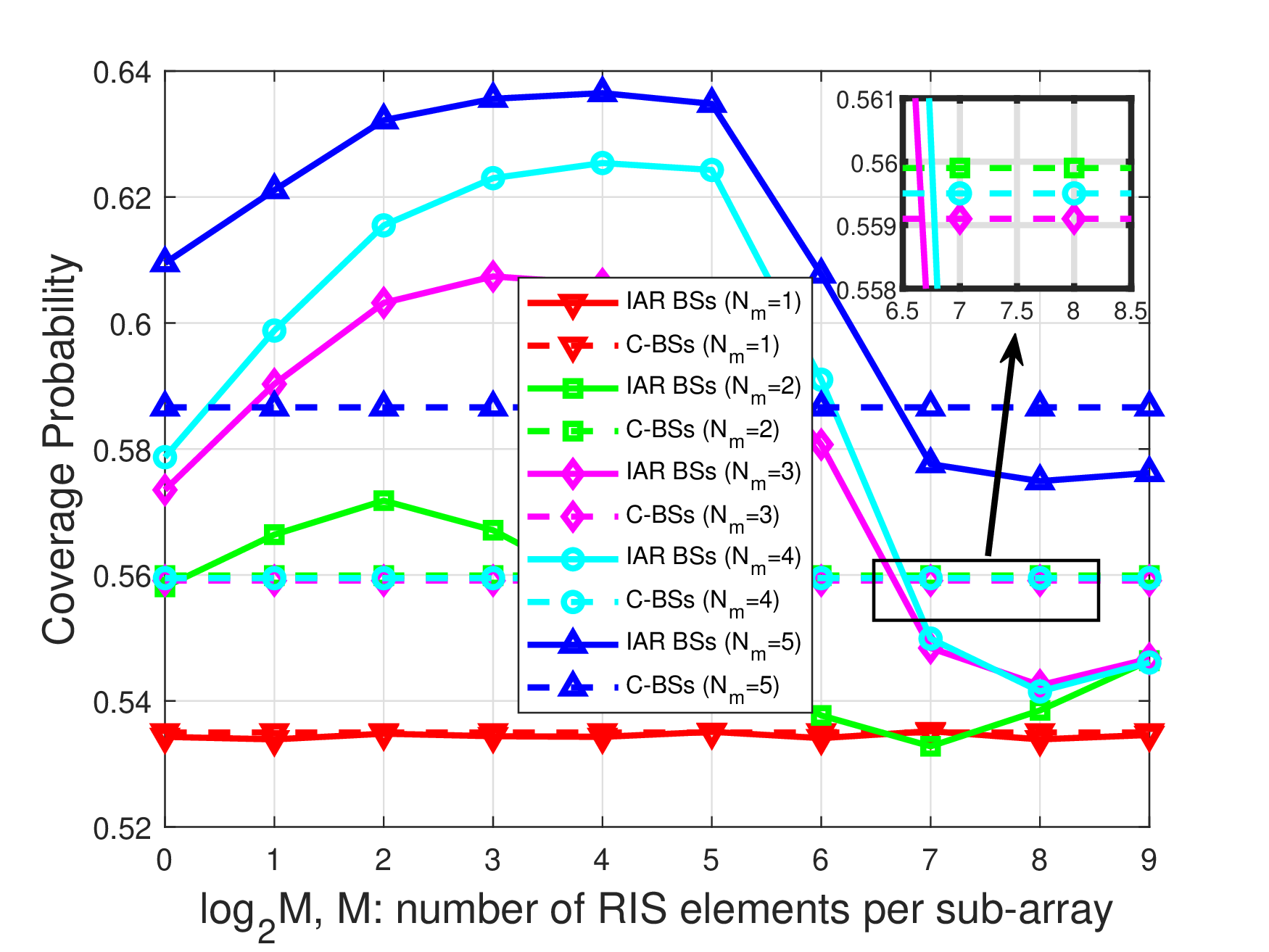}
    \caption{Coverage probability vs $\rm log_2M$ in a cell-free network considering the fixed sub-carrier scheme.}
    \label{fig:covpro_RA_cellfree}
\end{figure}
First, we consider a cell-free network under the fixed sub-carrier scheme. We count the number of IoT devices whose maximum SINR at the connected BSs is larger than the decoding threshold ${\rm SINR_{th}}$, \ie selection combining, and obtain the coverage probability through dividing it by the total number of IoT devices. As shown in Fig. \ref{fig:covpro_RA_cellfree}, when $M$ is less than $4$, the coverage probabilities generally increase because the reflections through RISs strengthen the signal power and support good enough diversity branches for devices. Different values of $N_m$ refer to different optimal designs of $M$. For example, when $N_m=2$, the optimal value of $M$ is 4. When $N_m=5$, the optimal value of $M$ is 16. This is because the reflection and amplification lead to a higher SINR in iLoS path than the dLoS path suffering from severe interference at the dLoS BS. Cell-free networks with IAR BSs in optimal RIS element allocation can outperform the networks with C-BSs. When $M>32$, the coverage probabilities generally decrease because the reflections are too strong, resulting in excessive interference to other devices. However, when $M$ is closer to $M_{tot}$, the performance slightly improves as the number of strong interfering iLoS paths decreases.  

\begin{figure}[ht]
    \centering
    \includegraphics[width=1\linewidth]{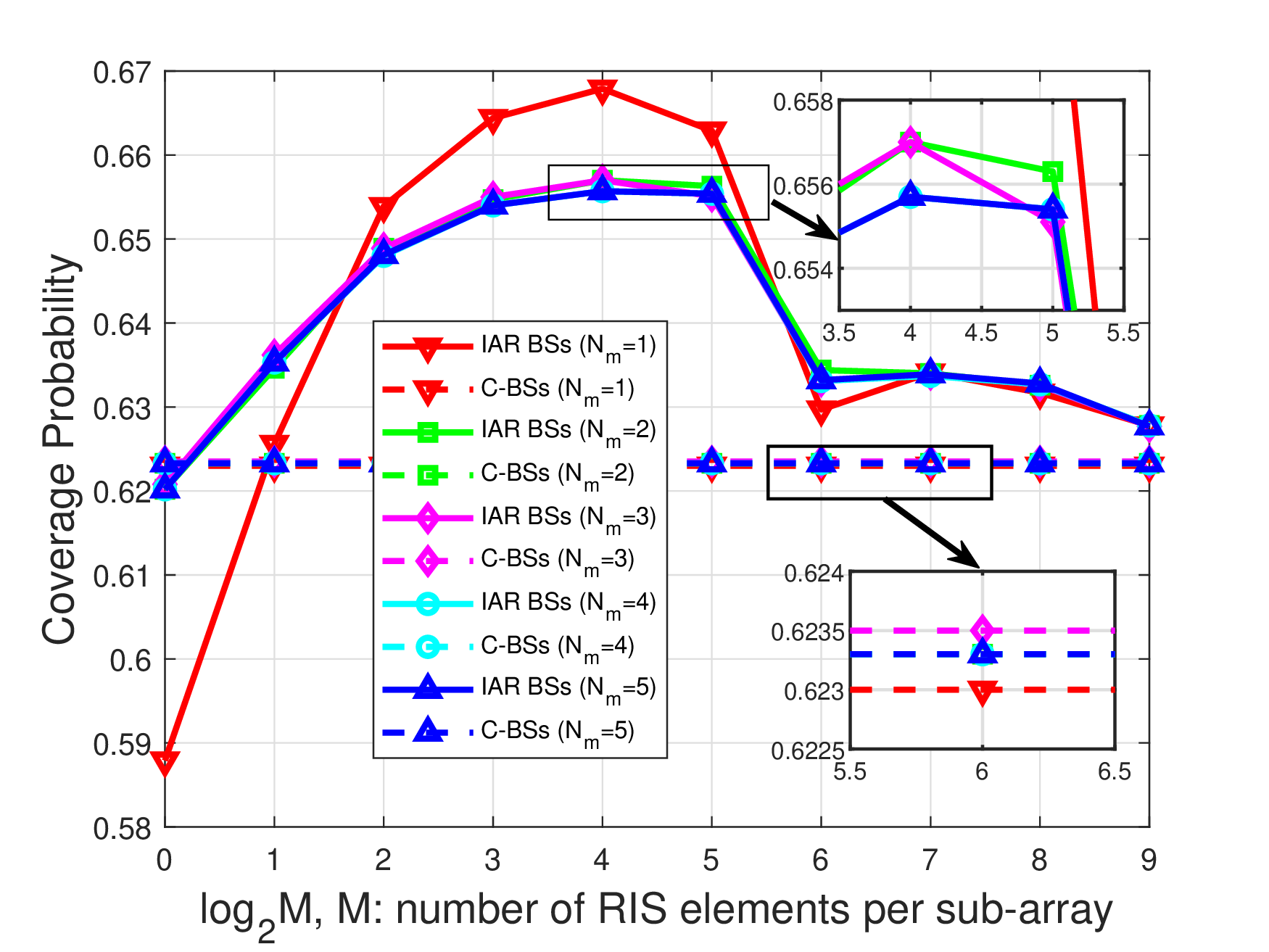}
    \caption{Coverage probability vs $\rm log_2M$ in a cell-free network considering the dynamic sub-carrier scheme.}
    \label{fig:covpro_AA_cellfree}
\end{figure}

\indent Next, as shown in Fig. \ref{fig:covpro_AA_cellfree}, we consider a cell-free network under the dynamic sub-carrier scheme. Because the dynamic sub-carrier scheme utilizes the idle resource blocks, the coverage probabilities with different values of $N_m$ and $M$ are generally better than that under the fixed sub-carrier scheme. Differently, when each resource block opens access to more IoT devices, the optimal performance decreases. This is because the dynamic sub-carrier scheme reduces the interference in each resource block, but opening access rights leads to higher interference again. When $M<16$, as the number of RIS elements used for each reflection increases, the iLoS links can be decoded more successfully as $M$ increases, so that coverage probabilities increase. When $M>16$, the reflection links can already be successfully decoded, the increase in $M$ will cause interference with others, and the decrease in the number of iLoS paths also makes some devices lose their service. These reasons lead to lower coverage performance than the case where $M=16$. When each IoT device operates only in its fixed sub-carrier, allowing multiple devices to share the same spectrum resource can enhance the link management capabilities of IAR. Conversely, when IoT devices can dynamically select the sub-carriers, each spectrum resource should not allow too many devices to access. Otherwise, devices will directly establish connections to dLoS BSs with a low quality of service, rather than utilize the relay scheme to obtain additional performance improvements.

\subsection{IAR-empowered cellular network}
\begin{figure}[ht]
    \centering
    \includegraphics[width=1\linewidth]{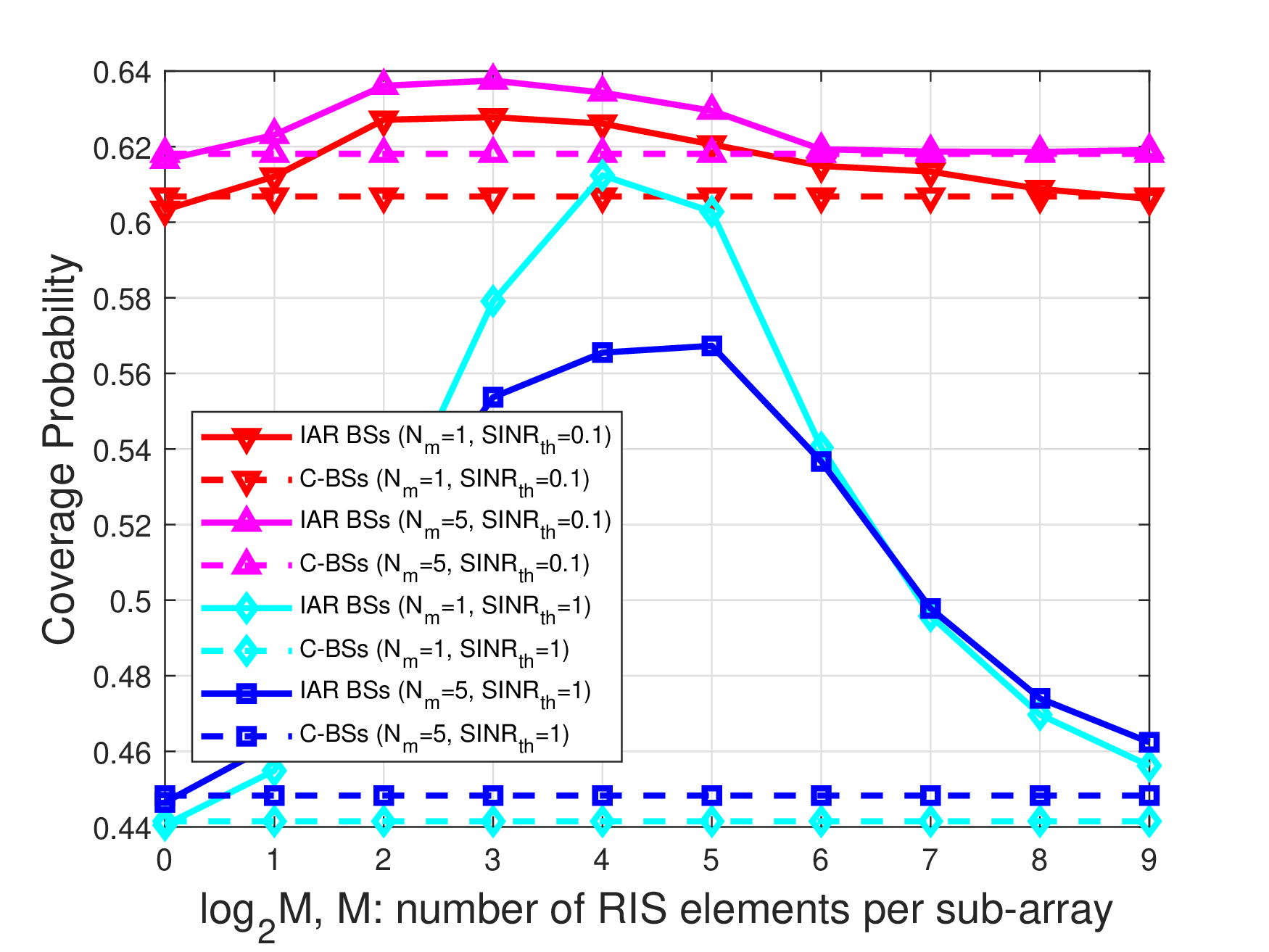}
    \caption{Coverage probability vs $\rm log2(M)$ in a cellular network considering the dynamic sub-carrier scheme.}
    \label{fig:covpro_AA_cellular}
\end{figure}
IAR BSs can not only enhance the cell-free services but also help offload the heavy tasks for BSs in cellular networks. To prove this, we take a cellular network with the dynamic sub-carrier scheme as an example. As shown in Fig. \ref{fig:covpro_AA_cellular}, the coverage probability can be improved by around $2\%$ when the SINR threshold is 0.1 and $N_m=1$. However, when the $\rm SINR_{th}=1$, the coverage probability can be improved by $18\%$ when $N_m=1$ and by $12\%$ when $N_m=5$. Therefore, the IAR BS makes it easier to satisfy the higher decoding SINR requirement than the conventional cellular network. Notice that such performance improvement is for the entire network when there are only 15 groups of co-sited RIS elements equipped, the hardware overhead of RISs is affordable. Similar to the cell-free case, the increase in $M$ when $M<16$ makes more iLoS paths successfully serve the devices, which also leads to high interference and resource competition to others when $M>16$. It is worth noting that under a low SINR decoding threshold $0.1$, increasing $N_m$ from $1$ to $5$ enables more devices to be successfully decoded, thereby achieving better coverage performance. However, when the SINR decoding threshold is set to $1$, the coverage enhancement achieved by IAR with $N_m=1$ outperforms that with $N_m=5$, since the spectral resources of idle BSs can be utilized more efficiently.
\\

It is worth noting that there exist many different access strategies and reflection strategies, which depend on different requirements of mobile networks. In this paper, we conduct Monte Carlo simulations to prove that such IAR technique can provide performance improvement for cellular and cell-free cases. Our simulation results provide guidance for the pre-allocation of RIS elements. In realistic cases, operators can use digital twin to simulate the traffic offloading via IAR and then optimize the RIS element allocation scheme. Dynamic RIS element allocation and phase reconfiguration can be developed, which may lead to greater complexity and latency. Using multiple-input multiple-output (MIMO) technique, non-orthogonal multiple access (NOMA) technique and adaptive resource management, the coverage performance can be further improved. Also, IAR BSs do not conflict with traditional RIS applications, such as creating first-hop LoS between more devices and BSs and enhancing the design of wave-domain transceivers. They can complement, integrate, or collaborate with each other. Next, we show more future opportunities and challenges of the application of the proposed IAR technique.

\section*{Opportunities and Challenges} \label{sec:challenge}
In this article, we introduced the concept of IAR and demonstrated its performance improvements for both traditional cellular and cell-free networks by supporting the collaboration between BSs. Furthermore, it can also be applied to more scenarios in future communication systems.

\textbf{Opportunity 1: IAR for cooperative communications.} In practical applications, IoT devices may have multiple available dLoS and iLoS paths, further enhancing the performance of cell-free networks. For example, when a device has only one dLoS, multi-hop reflections and multi-directional reflections are expected to be considered. In the realistic system with massive IoT devices, more advanced access strategies are expected to be discussed, considering user fairness and overall performance improvement.

\textbf{Opportunity 2: IAR for vertical heterogeneous networks.} In next-generation wireless networks, IAR technology can be used not only to upgrade terrestrial infrastructure and assist in wave-domain cooperation between BSs, but also to upgrade non-terrestrial infrastructure like drones, high altitude platform stations, and satellites. For instance, a drone can collect and decode device information, or reflect the communication requests to BSs. The tilt angle of RISs can be also optimized to enhance reflection links between the device and the iLoS BSs. This technology can promote wave-domain cooperation between terrestrial and non-terrestrial networks, resulting in a more reliable vertical heterogeneous network.

\textbf{Opportunity 3: IAR with sensing, powering and backhauling:} IAR technique not only benefits the communications, but also provides the opportunity for sensing, powering and backhauling. For example, when the spectrum resources of dLoS BS are busy, co-sited RISs can provide nearby devices with links to other BSs. These links can be used not only for communication, but also for multi-hop sensing and powering when the energy supply of each BS is constrained. The IAR functions provides the hardware base to realize software-defined radio (SDR), smart radio environment and 5G new radio (NR) \cite{renzo2019smart}.

\textbf{Opportunity 4: IAR with computing and learning.} IAR solution also provides the opportunity for joint edge computing and cloud computing. The edge nodes can decode the information from interested devices and reflect other signals directly to the core network. The integration of computing and learning can help improve the security and efficiency of IAR solutions \cite{mao2017survey,challita2020machine}. Further, artificial intelligence is expected to be used in managing the IAR system.\\ 
 
However, the applications of IAR techniques still face many challenges. We also discuss our concerns that may affect the realization of IAR.

\textbf{Challenge 1.} \textbf{Realization of hardware upgrade.} Currently, many multifunctional RISs have been proposed to achieve energy-efficient signal forwarding. However, RIS hardware remains difficult to deploy on a large scale. Although IAR architecture only requires the deployment of RIS components on BSs rather than all blockages, its affordability and return on investment remain important considerations for practical applications.

\textbf{Challenge 2.} \textbf{More complex radio environment.} Similar to the case where RISs are deployed on blockages, the IAR BSs may make the electromagnetic environment more complex. It may bring interference and frequency resource competition to the existing radio environment. 

\textbf{Challenge 3.} \textbf{Adaptive resource management.} The deployment of BS networks not only needs to consider stable connection requirements, but also adapt to dynamic traffic changes. For example, if a region has a high throughput requirement, the devices inside it need to use a multi-hop relay to access the BSs outside the region. Because the traffic and location of each device will change, the entire network should adopt a dynamic resource management and IAR strategy. Similar to traditional RIS solutions, the control plane of IAR BSs must simultaneously plan spectrum resources, user beams, and RIS elements and phase control \cite{10802983}. RISs are expected to be integrated into the design of new medium access control (MAC) layer protocols to solve the random access problem \cite{10109667}. In scenarios with highly dynamic traffic variations, the burden of RIS reconfiguration cannot be neglected.

\textbf{Challenge 4.} \textbf{Realization of cooperative system.} In IAR systems, each BS needs to synchronize the access requests with its neighboring BSs to determine the access path for each specific device. Due to the large number of RIS components, the whole network requires more computation resources to support RF resource allocation and phase alignment of RIS resources. Furthermore, while more devices are accessed in the context of cooperative communication, advanced cooperation schemes with lower signaling latency and overhead are important to be investigated in future work.

\section*{Conclusion} \label{sec:conclusion}
\indent In this paper, we introduce the concept of integrated access and relay (IAR) that can provide opportunities for advanced functions of BSs. The use of reconfigurable intelligent surfaces (RISs) can upgrade conventional BSs to IAR BSs. In cellular networks, IAR BSs can help offload the tasks surrounding the busy BSs to the idle BSs. In cell-free networks, IAR BSs provide more iLoS paths for IoT devices. We show that the IAR architecture can improve the coverage performance considering the fixed sub-carrier scheme and dynamic sub-carrier scheme. We discuss the relationship between the number of RIS elements per reflection and the coverage probability. We also introduce opportunities and challenges for the realization of IAR architecture, including adaptive resource management and inter-BS collaboration for more complex radio environments.

\ifCLASSOPTIONcaptionsoff
  \newpage
\fi

\bibliographystyle{IEEEtran}
\bibliography{ref}


\end{document}